\documentclass[aps,pra,preprint,groupedaddress]{revtex4-2}
\usepackage{amsmath}
\usepackage{graphicx}
\usepackage{ulem}
\usepackage[thicklines]{cancel}

\begin{document}


\title{On physical origin of Berry connection of free relativistic electrons}




\author{Chun-Fang Li}
\email[E-mail address: ]{cfli@shu.edu.cn}
\affiliation{Department of Physics, Shanghai University, 99 Shangda Road, 200444 Shanghai, China}



\begin{abstract}

No consensus has been reached on the appropriate operators for the spin and orbital angular momentum of free relativistic electrons. But there is a tendency to believe the existence of a spin-orbit coupling through the Berry connection. Here we show that this is not the case. 
Because the positive-energy solution of the Dirac equation is associated with the state of the electron, it is reasonable to postulate that only when projected onto the positive-energy subspace can observable operators on the Hilbert space of the Dirac equation represent physical quantities of the electron. 
On this basis, we find, after mapping the projected operators onto an irreducible representation, that the Berry connection describes the position of a quantum reference point with respect to the laboratory reference frame, relative to which the position of the electron is canonical. The orbital angular momentum of the quantum reference point about the origin of the laboratory reference frame is expressed in terms of the Berry connection. It combines with the canonical orbital angular momentum of the electron about the quantum reference point to give the orbital angular momentum of the electron about the origin of the laboratory reference frame.

\end{abstract}



\maketitle


\newpage

\section{Introduction}
It is often stated \cite{Bial-B, Barn} on the basis of the Dirac equation \cite{Dirac, Thal} that the spin and orbital angular momentum (OAM) of a free relativistic electron are not separately conservative. This is ascribed \cite{Blio-DN11, Blio-DN17} to the existence of the spin-orbit coupling (SOC) described by a gauge potential, the Berry connection \cite{Math, Blio, Bera-M, Chan-N, Chuu-CN}. No one seems to have realized that there is not any mechanism to render the spin and OAM of the electron coupled together. 
Here it is shown, through a proper interpretation of the spin and OAM operators in the Dirac theory, that the spin and OAM of the electron are separately conservative. The SOC does not exist. 
The presence of the non-Abelian Berry connection in this case reflects the fact that the canonical position \cite{Chan-N, Chuu-CN} of a free relativistic electron is not the same as its position with respect to the laboratory reference frame (LRF) \cite{Clau}, contrary to what was assumed in Refs.~\cite{Blio-DN11, Blio-DN17}.

As is taught in textbooks \cite{Rose}, the free Dirac equation is able to predict the existence of the spin of the Dirac particle because the OAM is found not to be conservative. The spin is such an angular momentum that is added to the OAM to give the conservative total angular momentum (TAM). The physics that requires the TAM of a free Dirac particle to be conservative is that no direction in space is preferred. However, as is known, the Dirac equation has both positive-energy and negative-energy solutions. The positive-energy solution is associated with the electron state and (the charge-conjugate transformation of) the negative-energy solution is associated with the positron state. Quantum mechanically, the so-called Dirac particle may be considered as an electron-positron system \cite{Frad-G, Barut-B} in the sense that the general solution of the Dirac equation can be divided into a superposition of positive-energy and negative-energy solutions.
Furthermore, according to von Neumann \cite{Neum}, an observable operator of a quantum system is defined on the Hilbert space spanned by the wave function of the system. This means that all the observable operators of the Dirac particle are defined on the whole Hilbert space of the Dirac equation.
Now that what is associated with the electron state is the position-energy solution, it is justified to postulate \cite{Pryce} that any observable operator of the electron is the projection of the corresponding observable operator of the Dirac particle onto the positive-energy subspace.
The problem is that so projected operator is mathematically reducible with the negative-energy subspace being its kernel. After mapping the projected operator onto the Hilbert space of an irreducible representation, one will get the corresponding irreducible operator.

With this approach, we find that the Berry connection of a free relativistic electron describes the position of a quantum reference point (QRP) with respect to the LRF, relative to which the position of the electron is canonical. 
By this it is meant that the position of the electron with respect to the LRF, referred to as the laboratory position of the electron, is the canonical position relative to the QRP plus the position of the QRP with respect to the LRF. It is no longer canonical.
Therefore, the OAM of the electron about the origin of the LRF, which is non-canonical, consists of two terms. One is the canonical OAM of the electron about the QRP. The other is the OAM of the QRP about the origin of the LRF. 
The spin of the electron also consists of two terms. One is represented by the well-known Pauli spin operator, which is canonical. The other depends on the Berry connection and is just exactly equal and opposite to the OAM of the QRP. The physics that determines the spin of the electron to have such a Berry-connection-dependent term is that the TAM of the electron should appear as the generator of the spatial rotation and hence satisfy the canonical commutation relation of angular momentum \cite{Vars-MK}. 
The findings presented here not only shed new light on the physical meaning of the angular momentum of the relativistic electron but will also deepen understanding of the physics of Berry phase \cite{Vand, Ciso-GF}.

\section{Conservation of the OAM and spin}
Let us start with discussing the conservation of the OAM and spin of free relativistic electrons. Denote by $\mathcal{H}^D$ the Hilbert space of the Dirac state vector $\psi_D$ that satisfies the free Dirac equation,
\begin{equation}\label{DE}
	i \hbar \frac{\partial \psi_D}{\partial t} =H_D \psi_D,
\end{equation}
where $H_D =c \mathbf{p} \cdot \boldsymbol{\alpha} +mc^2 \beta$ 
is the free Dirac Hamiltonian, 
\[\boldsymbol{\alpha} =\begin{pmatrix}
		                  0                   & \boldsymbol{\sigma} \\
		                  \boldsymbol{\sigma} & 0
	                   \end{pmatrix},                                 \quad
\beta= \begin{pmatrix}
		    1 & 0  \\
		    0 & -1
	    \end{pmatrix}         
\] 
are the $4 \times 4$ Dirac matrices in the Dirac representation \cite{Thal}, $\boldsymbol{\sigma}$ are the usual Pauli matrices, and $\mathbf p$ is the momentum.
It is the direct sum of the positive-energy and negative-energy subspaces. 
As mentioned above, an observable operator $O_D$ on the whole Hilbert space $\mathcal{H}^D$ represents a physical quantity of the Dirac particle.
Only when projected onto the positive-energy subspace $\mathcal{H}^D_e$, referred to as the electron subspace, can it serve as the operator for the corresponding quantity of the electron. 
So projected operator is mathematically defined as~\cite{Pryce}
\begin{equation}\label{Oe}
	O_e =\Lambda O_D \Lambda,
\end{equation}
where 
$\Lambda=\frac{1}{2}\left(1+\frac{H_D}{E}\right)$
is the projection operator from $\mathcal{H}^D$ onto $\mathcal{H}^D_e$, 
$E=(m^2 c^4 +c^2 p^2)^{1/2}$ is the energy of the electron, and $p=|\mathbf{p}|$.
As an example, the Hamiltonian of the free electron is given by
\begin{equation*}
	H_e =\Lambda H_D \Lambda =E \Lambda.
\end{equation*}

The Dirac position operator on $\mathcal{H}^D$ is
$\mathbf{x}_D =i\hbar \nabla_p$
in the momentum space, where $\nabla_p$ stands for the gradient operator with respect to the momentum. According to definition (\ref{Oe}), the position operator of the free electron on $\mathcal{H}^D_e$ is given by $\mathbf{x}_e =\Lambda \mathbf{x}_D \Lambda$, which is calculated to be
\begin{equation}\label{xe}
	\mathbf{x}_e =\left[\mathbf{x}_D -\frac{i\hbar c}{2E}
	                    \left(\boldsymbol{\alpha}-\frac{c \mathbf{p}}{H_D}\right)\right] \Lambda.
\end{equation}
Therefore, the operator for the OAM of the free electron on $\mathcal{H}^D_e$, which is defined as
$\mathbf{L}_e =\Lambda \mathbf{L}_D \Lambda$,
reads
\begin{equation}\label{Le}
	\mathbf{L}_e =\mathbf{x}_e \times \mathbf{p}
	=\left(\mathbf{L}_D -\frac{i\hbar c}{2E} \boldsymbol{\alpha} \times \mathbf{p}\right) \Lambda,
\end{equation}
where $\mathbf{L}_D =\mathbf{x}_D \times \mathbf{p}$
is the Dirac OAM operator on $\mathcal{H}^D$. It is conservative, obeying
$[\mathbf{L}_e, H_e]=0$.
In addition, since the Dirac spin operator on $\mathcal{H}^D$ is 
$
\mathbf{S}_D =\frac{\hbar}{2} \begin{pmatrix}
	                              \boldsymbol{\sigma} & 0 \\
	                              0                   & \boldsymbol{\sigma}
                              \end{pmatrix}
$,
the spin operator of the free electron on $\mathcal{H}^D_e$ is given by 
\begin{equation}\label{Se}
	\mathbf{S}_e =\Lambda \mathbf{S}_D \Lambda
	             =\left(\mathbf{S}_D +\frac{i\hbar c}{2E} \boldsymbol{\alpha} \times \mathbf{p}\right) \Lambda.
\end{equation}
It is also conservative, obeying 
$[\mathbf{S}_e, H_e]=0$.
From Eqs.~(\ref{Le}) and (\ref{Se}) one sees that the operator for the TAM of the electron on $\mathcal{H}^D_e$ takes the form
\[\mathbf{J}_e =\mathbf{L}_e + \mathbf{S}_e =(\mathbf{L}_D +\mathbf{S}_D) \Lambda.\]
This is nothing but the projection of the Dirac TAM operator
$\mathbf{J}_D =\mathbf{L}_D +\mathbf{S}_D$
onto $\mathcal{H}^D_e$.
Indeed, due to the fact that $\mathbf{J}_D$ commutes with $H_D$ and hence with $\Lambda$, one has 
$\Lambda \mathbf{J}_D \Lambda =\mathbf{J}_D \Lambda$
for the projection of $\mathbf{J}_D$ onto $\mathcal{H}^D_e$.
This confirms that the position and spin operators of the electron on $\mathcal{H}^D_e$ are truly given by Eqs.~(\ref{xe}) and (\ref{Se}), respectively.  

As is known, the Dirac OAM operator $\mathbf{L}_D$ on $\mathcal{H}^D$ is canonical, satisfying the canonical commutation relation of angular momentum,
\[\mathbf{L}_D \times \mathbf{L}_D =i\hbar \mathbf{L}_D.\]
This is because the Dirac position operator $\mathbf{x}_D$ on $\mathcal{H}^D$ is canonical, satisfying
\begin{equation*}
	\mathbf{x}_D \times \mathbf{x}_D =0.
\end{equation*}
But from this equation it is not difficult to find that the operator (\ref{xe}) for the electron position is no longer canonical \cite{Bacry}, satisfying the following commutation relation, 
\begin{equation}\label{CR-xe}
	\mathbf{x}_e \times \mathbf{x}_e =-\frac{i\hbar c^2}{E^2} \mathbf{S}_e.
\end{equation}
As a consequence, the OAM of the electron represented by operator (\ref{Le}) is not canonical. Of course, Eq.~(\ref{Se}) means that the spin of the electron is dependent on its momentum, or equivalently, on its velocity. In fact, it can be readily checked that
\begin{equation}\label{SoSO} 
	\mathbf{S}_e \cdot \mathbf{S}_e 
	=\frac{\hbar^2}{4} \left[1+2\left(\frac{mc^2}{E}\right)^2\right] \Lambda
	=\frac{\hbar^2}{4} \left(1+\frac{2}{\gamma^2}\right) \Lambda,
\end{equation}
where 
$\gamma=(1-v^2/c^2)^{-1/2}$
and $v$ denotes the velocity of the electron.
Nevertheless, Eqs.~(\ref{Le}) and (\ref{Se}) do not tell how and why the OAM and spin of the electron depend on the Berry connection. The point is that both the position operator $\mathbf{x}_e$ and the spin operator $\mathbf{S}_e$ are mathematically reducible as they all commute with the projection operator $\Lambda$ \cite{Kadi-R}. In fact, as was shown by Foldy and Wouthuysen \cite{Foldy-W}, the wave function of the free electron is essentially a two-component spinor. That is to say, the irreducible representation of the free electron is merely a two-component representation.
Let us make use of the Foldy-Wouthuysen (FW) transformation to deduce a transformation that maps the position and spin operators onto the Hilbert space of an irreducible representation.

\section{Quasi unitary transformation}
As mentioned above, it is the wave function in the electron subspace $\mathcal{H}^D_e$ that describes the state of the electron. That is to say, the electron wave function in the Dirac representation, denoted by $\psi_{D,e}$, is the projection of the Dirac wave function $\psi_D$ onto $\mathcal{H}^D_e$,
\begin{equation}\label{psie}
	\psi_{D,e} =\Lambda \psi_D.
\end{equation}
The FW transformation, which reads \cite{Foldy-W}
\begin{equation}\label{U} 
	U =\frac{E +\beta H_D }{[2 E (E +mc^2)]^{1/2}},
\end{equation}
maps $\psi_D$ into
\begin{equation}\label{psiFW}
	\psi_{FW} =U \psi_D
\end{equation}
in the FW representation. Accordingly, it maps $\psi_{D,e}$ into
\begin{equation}\label{FWT}
	\psi_{FW,e} =U \psi_{D,e},
\end{equation}
the wave function of the electron in the FW representation. Upon substituting Eq. (\ref{psie}) and considering the inverse of Eq. (\ref{psiFW}), one has
\begin{equation*}
	\psi_{FW,e} =U \Lambda U^\dag \psi_{FW} =\frac{1}{2} (1+\beta) \psi_{FW}.
\end{equation*}
This shows that $\psi_{FW,e}$ has a vanishing lower part \cite{Foldy-W}, assuming the form of
\begin{equation}\label{EWF-FW}
	\psi_{FW,e} =\begin{pmatrix}
		             \psi \\
		             0
	             \end{pmatrix}.
\end{equation}
Here the upper part $\psi$ is just the two-component wave function of the electron. The Hilbert space of $\psi$, denoted by $\mathcal{H}$, constitutes an irreducible representation of the electron. 
Substituting expressions (\ref{U}) and (\ref{EWF-FW}) into Eq.~(\ref{FWT}), one finally gets
\begin{equation}\label{QUT}
	\psi =Q \psi_{D,e},
\end{equation}
where $Q$ is a $2 \times 4$ matrix, given by
\begin{equation*}
	Q =\frac{1}{[2E(E+mc^2)]^{1/2}}
	   \begin{pmatrix}
		   E+mc^2 & c\mathbf{p} \cdot \boldsymbol{\sigma}
	   \end{pmatrix},
\end{equation*}
which comes from the $2 \times 4$ upper sector of the matrix $U$. On the other hand, the inverse of transformation (\ref{FWT}) reads
\begin{equation*}
	\psi_{D,e} =U^\dag \psi_{FW,e},
\end{equation*}
from which one obtains, upon considering expressions (\ref{U}) and (\ref{EWF-FW}),
\begin{equation}\label{IQUT}
	\psi_{D,e} =Q^\dag \psi.
\end{equation}

Equations (\ref{QUT}) and (\ref{IQUT}) demonstrate that $Q$ is just the transformation to connect the electron subspace $\mathcal{H}^D_e$ of the Dirac representation with the Hilbert space $\mathcal{H}$ of the irreducible representation. Upon substituting Eq.~(\ref{IQUT}) into Eq.~(\ref{QUT}), one finds
\begin{equation}\label{unitarity1}
	Q Q^\dag =I_2,
\end{equation}
where $I_2$ is the $2 \times 2$ unit matrix. Furthermore, substituting (\ref{QUT}) into (\ref{IQUT}) gives
$Q^\dag Q \psi_{D,e} =\psi_{D,e}$,
which means that on the electron subspace $\mathcal{H}^D_e$, one must have
\begin{equation}\label{unitarity2}
	Q^\dag Q =I_4,
\end{equation}
where $I_4$ is the $4 \times 4$ unit matrix.
Mathematically speaking, Eqs.~(\ref{unitarity1}) and (\ref{unitarity2}) demonstrate that the transformation to connect $\mathcal{H}^D_e$ and $\mathcal{H}$ is a quasi unitary transformation \cite{Golub-L}. $Q^\dag$ is the pseudo inverse of $Q$ and vice versa. 
On this basis, the electron operator (\ref{Oe}) in the Dirac representation is mapped by $Q$ into
\begin{equation*}
	O_{ir} =Q O_e Q^\dag =Q \Lambda O_D \Lambda Q^\dag
\end{equation*}
in the irreducible representation. Since
$Q \Lambda =Q$
as can be readily checked, the above equation reduces to
\begin{equation}\label{O}
	O_{ir} =Q O_D Q^\dag.
\end{equation}
It shows that the transformation $Q$ enables to map the observable operator $O_D$ on $\mathcal{H}^D$ directly into the corresponding operator of the electron on $\mathcal H$.
For instance, the Hamiltonian of the electron in the irreducible representation is given by
\begin{equation}\label{H}
	H_{ir} =Q H_D Q^\dag =E
\end{equation}
as is expected. 

According to Eq.~(\ref{O}), the position operator of the electron in the irreducible representation is given by
\begin{equation}\label{x}
	\mathbf{x}_{ir} =Q \mathbf{x}_D Q^\dag 
	=\mathbf{q} +\mathbf{R},
\end{equation}
where $\mathbf{q} =i\hbar \nabla_p$ and
\begin{equation}\label{R} 
	\mathbf{R} =i\hbar Q \nabla_p Q^\dag
	=-\frac{\hbar c^2 \boldsymbol{\sigma} \times \mathbf{p}}{2E(E+mc^2)}.
\end{equation}
It is worth noting that the operator $\mathbf{q}$ here should not be confused with $\mathbf{x}_D$ in that they are defined on different Hilbert spaces. The former is defined on $\mathcal{H}$ and the latter on $\mathcal{H}^D$.
From Eq.~(\ref{x}) it follows that the operator for the OAM of the electron in the irreducible representation takes the form
\begin{equation}\label{L}
	\mathbf{L}_{ir} =\mathbf{x}_{ir} \times \mathbf{p}
	=\mathbf{L}_{can} +\mathbf{R} \times \mathbf{p},
\end{equation}
where
$\mathbf{L}_{can} =\mathbf{q} \times \mathbf{p}$.
Correspondingly, the spin operator of the electron in the irreducible representation is given by
\begin{equation*}
	\mathbf{S}_{ir} =Q \mathbf{S}_D Q^\dag 
	=\frac{\hbar}{2} \boldsymbol{\sigma}
	+\frac{\hbar c^2 (\boldsymbol{\sigma} \times \mathbf{p}) \times \mathbf{p}}{2E(E +mc^2)},
\end{equation*}
which can be rewritten in terms of $\mathbf{R}$ as 
\begin{equation}\label{S-R}
	\mathbf{S}_{ir} =\frac{\hbar}{2} \boldsymbol{\sigma} -\mathbf{R} \times \mathbf{p}.
\end{equation}
The second term here is equal and opposite to the second term of (\ref{L}). As a result, the operator for the TAM of the electron reads
\begin{equation}\label{J}
	\mathbf{J}_{ir} =\mathbf{L}_{ir} +\mathbf{S}_{ir}
	                =\mathbf{L}_{can} +\frac{\hbar}{2} \boldsymbol{\sigma}
\end{equation}
in the irreducible representation. It is no doubt that both the OAM (\ref{L}) and the spin (\ref{S-R}) are conservative.

The vector operator $\mathbf R$ in (\ref{R}) is nothing but the Berry connection of a free electron \cite{Math, Blio, Bera-M, Chan-N, Chuu-CN}. 
This is to say that expressions (\ref{L}) and (\ref{S-R}) manifest how the operators for the OAM and spin of the electron depend on the Berry connection. To understand further why they do so, we need to explain the physical meaning of the Berry connection.

\section{Physical meaning of the Berry connection}
It is seen from Eq.~(\ref{x}) that the Berry connection connects two different operators $\mathbf{x}_{ir}$ and $\mathbf{q}$. To explain the meaning of the Berry connection, the key point is to figure out what these two operators represent.
According to Dirac \cite{Dirac, Barut-M}, the Dirac operator $\mathbf{x}_D$ on the Hilbert space $\mathcal{H}^D$ represents the laboratory position \cite{Clau} of the Dirac particle. For this reason, it is proper to interpret $\mathbf{x}_{ir}$ on the Hilbert space $\mathcal{H}$ as the operator for the laboratory position of the electron.
Meanwhile, like $\mathbf{x}_D$ on $\mathcal{H}^D$, the Cartesian components of $\mathbf{q}$ on $\mathcal{H}$ should commute,
\begin{equation}\label{CR-r}
	\mathbf{q} \times \mathbf{q} =0.
\end{equation}
In fact, according to Eq.~(\ref{O}), the operator for the TAM of the electron on $\mathcal{H}$ is connected with $\mathbf{J}_D$ on $\mathcal{H}^D$ by
\begin{equation*}
	\mathbf{J}_{ir} =Q \mathbf{J}_D Q^\dag.
\end{equation*}
Since $\mathbf{J}_D$ leaves the electron subspace $\mathcal{H}^D_e$ invariant, upon using the commutation relation
$\mathbf{J}_D \times \mathbf{J}_D =i\hbar \mathbf{J}_D$
and taking Eq.~(\ref{unitarity2}) into consideration, one has 
\begin{equation}\label{CR-J}
	\mathbf{J}_{ir} \times \mathbf{J}_{ir} = i\hbar \mathbf{J}_{ir}.
\end{equation}
This in turn requires $\mathbf{q}$ to satisfy the commutation relation (\ref{CR-r}) by virtue of expression (\ref{J}). In other words, $\mathbf{q}$ represents the canonical position of the electron \cite{Chan-N, Chuu-CN}. 
Apparently, the canonical position is not the same as the laboratory position $\mathbf{x}_{ir}$ \cite{Clau}. By analogy with classical mechanics \cite{Gold}, one can see from Eq.~(\ref{x}) that it is the position of the electron relative to a reference point. The position of the reference point with respect to the LRF arises as the Berry connection $\mathbf R$. 

It is known that the Berry connection is non-Abelian \cite{Math, Blio, Bera-M, Chan-N, Chuu-CN}. This means that contrary to what we are familiar with in classical mechanics, here the reference point for the canonical position in quantum mechanics is nonlocal in the sense that its Cartesian coordinates are noncommutative \cite{Bacry},
\begin{equation}\label{CR-b}
	\mathbf{R} \times \mathbf{R} =\frac{i\hbar c^4 (\mathbf{S}_{ir} \cdot \mathbf{p})\mathbf{p}}{E^2 (E+mc^2)^2}.
\end{equation}
Even so, the QRP represented by $\mathbf R$ is a constant of motion, obeying
\begin{equation*}
	[\mathbf{R},H_{ir}]=[\mathbf{R},E]=0	
\end{equation*}
by virtue of Eq.~(\ref{H}). Consequently, the velocity of the free electron is given by
\begin{equation*}
	\frac{d \mathbf{x}_{ir}}{dt}
   =\frac{1}{i\hbar}[\mathbf{x}_{ir},H_{ir}]=\frac{1}{i\hbar}[\mathbf{q},E]=\frac{c^2 \mathbf{p}}{E},
\end{equation*}
as is expected. 
In a word, the Berry connection $\mathbf{R}$ serves as the operator for the position of the QRP with respect to the LRF, relative to which the electron position $\mathbf{q}$ is canonical. In view of this, Eq.~(\ref{x}) shows that the laboratory position of the electron is the canonical position of it relative to the QRP plus the position of the QRP with respect to the LRF.

It is now clear that the second term on the right side of Eq.~(\ref{L}), the Berry-connection-dependent term, represents the OAM of the QRP about the origin of the LRF. It combines with the first term, the canonical OAM of the electron about the QRP, to give the total OAM of the electron, the OAM about the origin of the LRF. That is to say, it itself is one part of the total OAM, having nothing to do with the so-called SOC. But due to the presence of the OAM of QRP, the total OAM is no longer canonical. It does not satisfy the canonical commutation relation of angular momentum,
\begin{equation*}
	\mathbf{L}_{ir} \times \mathbf{L}_{ir} \neq i\hbar \mathbf{L}_{ir}.
\end{equation*}
This can also be understood from the fact that the presence of the Berry connection makes the laboratory position $\mathbf{x}_{ir}$ non-canonical. Indeed, according to Eqs.~(\ref{CR-r}) and (\ref{CR-b}), the commutation relation of $\mathbf{x}_{ir}$ takes the form of
\begin{equation}\label{CR-x}
	\mathbf{x}_{ir} \times \mathbf{x}_{ir} =i\hbar \mathbf{B} =-\frac{i\hbar c^2}{E^2} \mathbf{S}_{ir},
\end{equation}
in agreement with Eq.~(\ref{CR-xe}), where
\begin{equation*}
	\mathbf{B} =\nabla_p \times \mathbf{R} +(i\hbar)^{-1}\mathbf{R} \times \mathbf{R}
	=-\frac{c^2}{E^2} \mathbf{S}_{ir}
\end{equation*}
is the Berry curvature \cite{Blio, Bera-M, Chan-N}. 

On the other hand, since there is no preferred direction in space for a free electron, the TAM of it should act as the generator of spatial rotation \cite{Vars-MK}. This requires that the spin of the electron, when added to the OAM (\ref{L}), should give a canonical TAM. The spin operator (\ref{S-R}) just meets such a requirement. Possessing the term that exactly cancels out the OAM of the QRF, it combines with the OAM operator (\ref{L}) to give the TAM operator (\ref{J}), which satisfies the canonical commutation relation (\ref{CR-J}). This is why the operator (\ref{S-R}) for the spin of the electron depends on the Berry connection.
The consequence is that the spin (\ref{S-R}) depends on the momentum or on the velocity. In fact, one can convert expression (\ref{S-R}) into
\begin{equation*}
	\mathbf{S}_{ir} =\frac{\hbar}{2 p^2} 
	\left[(\boldsymbol{\sigma} \cdot \mathbf{p}) \mathbf{p} 
	-\frac{1}{\gamma} (\boldsymbol{\sigma} \times \mathbf{p}) \times \mathbf{p}\right],
\end{equation*}
manifesting that the velocity diminishes only the transverse component of the spin. 
A straightforward calculation gives 
\begin{equation*}
	\mathbf{S}_{ir}^2 =\frac{\hbar^2}{4} \left(1+\frac{2}{\gamma^2}\right),
\end{equation*}
in agreement with Eq.~(\ref{SoSO}). The Berry-connection dependent term is a relativistic correction. Letting
$\mathbf{S}_{BCD} =-\mathbf{R} \times \mathbf{p}$, one has
$\mathbf{S}_{BCD}^2 =\frac{\hbar^2}{2} \left(1-\frac{mc^2}{E}\right)^2$.

To be honest, Bliokh \textit{et al.} \cite{Blio-DN11, Blio-DN17} made use of the same equation as (\ref{x}) to discuss the properties of the OAM and spin of the electron. But, unfortunately, they looked upon it as a new choice of the electron position operator \cite{Thal, Pryce, Cele-KT}. The reason is, as was indicated by their Eqs.~(4), (13), and (14) in Ref. \cite{Blio-DN17}, that they confused the canonical-position opertor $\mathbf{q}$ in Eq.~(\ref{x}) on the Hilbert space $\mathcal{H}$ with the Dirac position operator $\mathbf{x}_D$ on $\mathcal{H}^D$. That is, they took the canonical OAM $\mathbf{L}_{can}$ as the OAM of the electron about the origin of the LRF. Under such a premise, they misinterpreted the second term of (\ref{L}), the OAM of the QRP, as arising from the SOC. 
Anyway, the Hamiltonian (\ref{H}) does not contain any term that describes an SOC effect.

\section{Concluding Remarks}

All the results presented here are obtained on the basis of the postulation expressed by Eq.~(\ref{Oe}). That is, the operator for an observable of the electron is the projection of the corresponding Dirac operator onto the electron subspace. From this point of view, one may say that the free Dirac equation (\ref{DE}) predicts not only the existence of the electron spin but also its dependence on the momentum as is shown by Eqs.~(\ref{Se}) or (\ref{S-R}).

It is seen from Eq.~(\ref{CR-xe}) or (\ref{CR-x}) that due to the existence of the spin, the laboratory position of a free electron is non-canonical, even in the non-relativistic limit $E \rightarrow mc^2$. This suggests, in accordance with the correspondence principle, that the Cartesian coordinates of the laboratory position of a free electron in relativistic quantum mechanics cannot be treated as independent degrees of freedom \cite{Clau}. The significance of this corollary deserves more attentions and further studies.





\end{document}